\documentclass[twocolumn]{aastex62}

\newcommand{\magphys}{\texttt{MAGPHYS}}
\newcommand{\magphysz}{\texttt{MAGPHYS+photo}-$z$}
\newcommand{\zphot}{$z_\mathrm{phot}$}
\newcommand{\photoz}{photo-$z$}
\newcommand{\zspec}{$z_\mathrm{spec}$}

\begin{document}

\title{Avoiding (photo-$z$) Catastrophe} %\footnote{Footnotes can be added to titles}}

\correspondingauthor{A. J. Battisti}
\email{andrew.battisti@uwa.edu.au}

\author[0000-0003-4569-2285]{A. J. Battisti}
\affil{International Centre for Radio Astronomy Research, University of Western Australia, 35 Stirling Hwy, Crawley, WA 6009, Australia}
\affil{Research School of Astronomy and Astrophysics, Australian National University, Cotter Road, Weston Creek, ACT 2611, Australia}

\author[0000-0001-9759-4797]{E. da Cunha}
\affil{International Centre for Radio Astronomy Research, University of Western Australia, 35 Stirling Hwy, Crawley, WA 6009, Australia}
%\email{elisabete.dacunha@uwa.edu.au}

\author[0000-0002-8412-7951]{S. Jin}
\affil{Cosmic Dawn Center (DAWN), Denmark}
\affiliation{DTU Space, Technical University of Denmark, Elektrovej 327, DK-
2800 Kgs. Lyngby, Denmark}
%\email{shuji@dtu.dk}

\author[0000-0001-6586-8845]{J. A. Hodge}
\affil{Leiden Observatory, Leiden University, NL-2300 RA Leiden, Netherlands}
%\email{hodge@strw.leidenuniv.nl}

%\collaboration{all}{The Terra Mater collaboration}

\begin{abstract}

Spectral modeling codes that estimate photometric redshifts (\photoz) are a powerful and often reliable method for determining redshifts of galaxies. However, there are notable instances where degeneracies in spectral energy distribution (SED) colors lead to `catastrophic' failures. We highlight the case of COSBO-7, a dusty, intermediate-$z$ galaxy that masqueraded as a high-$z$ source, because it demonstrates a unique scenario where \photoz\ codes run into issues despite extensive multi-wavelength photometry. We advocate that \photoz\ fitting should aim to: (1) use the entire available SED (UV--radio) whenever possible to help break color degeneracies, (2) allow flexible dust attenuation prescriptions, both in terms of the attenuation curve slope and a varying 2175\AA\ absorption feature, and (3) implement uncertainty floors to account for limitations in spectral models and also on the photometry itself.

% ApJ, ApJL, AJ, and PSJ all have a 250 word limit for the abstract. The limit is 150 for RNAAS manuscripts.
\end{abstract}

%\keywords{\uat{Galaxies}{573} --- \uat{High-redshift galaxies}{734} --- \uat{Interstellar dust extinction}{837} --- \uat{Spectral energy distribution}{2129} --- \uat{Astronomy data analysis}{1858}}


\section{Introduction} 
%The James Webb Space Telescope (JWST) is delivering on expectations to detect the first galaxies \citep[$z_\mathrm{spec}\gtrsim13$; e.g.,][]{curtis-lake23}. 
Galaxies at high-$z$ are often identified as candidates through photometric redshifts (\photoz\ or \zphot) derived from their spectral energy distribution \cite[SED; ][]{salvato19}\footnote{We ignore machine-learning \zphot\ methods.}. %, typically using a moderate number of observer-frame optical and near-IR filters. 
However, \photoz\ codes can have instances of `catastrophic' failure, often defined as $\vert z_\mathrm{phot}-z_\mathrm{spec}\vert/(1+z_\mathrm{spec})>0.15$ \citep{hildebrandt12}, with low-$z$, dusty galaxies masquerading as high-$z$ candidates being a common case \cite[e.g.,][]{zavala23, jin24}. Given the investment of telescope time used for redshift confirmation, it is important to thoroughly test \photoz\ solutions before pursuing those avenues, even in instances where high quality JWST data are available.

\section{The Case of COSBO-7} 
COSBO-7 is a strongly-lensed galaxy in COSMOS/PRIMER \citep{scoville07, dunlop21}, % with extensive multi-wavelength photometry \citep[e.g.,][]{jin18, weaver22}. 
first presented in the COSBO survey \citep{bertoldi07} with a IR+radio-based \photoz\ of $z_\mathrm{phot}=3.1^{+0.5}_{-0.4}$. %, and later with $z_\mathrm{phot}=3.4\pm0.4$ by \cite{pearson24}. 
Later, \cite{ling24} found that \photoz\ based on 10 bands of JWST/NIRCam+MIRI across numerous codes favored $z_\mathrm{phot}\gtrsim7$ and also presented an ALMA Band~3 line detection that could be CO(7–6) at $z=7.455$. Unexpectedly, the low-$z$ solution was realised to be correct after detections of CO(7–6) and [CI](2–1) at $z=2.625$ from ALMA Band~6 (instead of [CII]; \citealt{jin24}), with the Band~3 line being CO(2–1). 
%Our intention with this Note is not to advocate for a particular \photoz\ code, but to highlight some of their limitations. 

We adopt photometry of COSBO-7 from \citet[][$0.9-18\mu$m from JWST/NIRCam+MIRI]{ling24} and \citet[][$24\mu\mathrm{m}-20$cm from multiple facilities]{jin24}. 
We use the \magphysz\ code \citep{battisti19}, and we refer readers to the aforementioned paper for details. Briefly, \magphysz\ is a modified version of \magphys\ \citep{dacunha08, dacunha15} that combines physically-motivated models for stellar and dust emission to fit ultraviolet-to-radio photometry and determine galaxy \zphot\ likelihoods and physical properties simultaneously. \magphysz\ assumes a \zphot\ prior that peaks at $z\sim2$ and decreases at both lower and higher redshift (similar to cosmic SFR density, \citealt{madau&dickinson14}). %, which is optimized for application to dusty star forming galaxies (SFGs). 
It assumes a 10\% uncertainty floor and does not perform any `tuning' to optimize residuals/zeropoints. %that occurs in some \photoz\ codes

%% The "ht!" tells LaTeX to put the figure "here" first, at the "top" next
%% and to override the normal way of calculating a float position.
%% The asterisk after "figure" tells the compiler to span multiple columns
%% if a two column style is selected.
\begin{figure*}[ht!]
%\plotone{COSBO7_compare-crop.pdf}
\includegraphics[width=\textwidth]{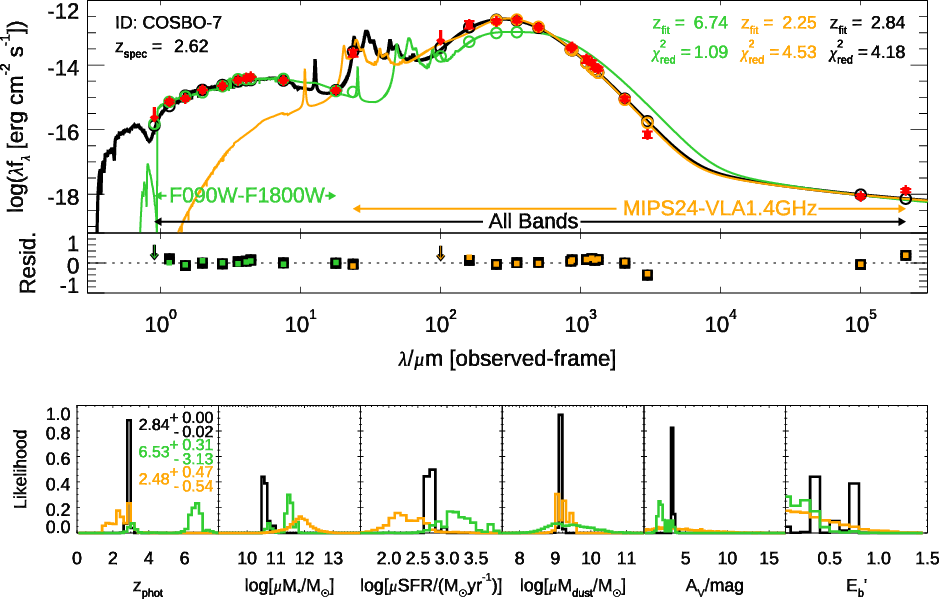}
\caption{\textit{Upper panel:} \magphysz\ best-fit solutions for three filtersets: F090W--F1800W (green), MIPS24--VLA1.4GHz (orange), and all filters (black). \textit{Lower panel:} Likelihood distributions for \zphot\ (median and 1$\sigma$ listed) and a few galaxy properties (some uncorrected for lensing magnification, $\mu$). $E_b'$ describes the 2175\AA\ feature strength \citep{battisti19}.
For JWST-only filters (green), there is a degeneracy where the F090W upper-limit can be interpreted as either a Lyman break or a moderate 2175\AA\ dust absorption feature, resulting in two \zphot\ peaks (first panel, second row). For filters longer than JWST (orange), the \zphot\ solution is very broad but single-peaked at low-$z$. Combining all filters (black) provides the tightest constraint on \zphot .
\label{fig:fig1}}
\end{figure*}

Figure~\ref{fig:fig1} presents the \magphysz\ results for COSBO-7 assuming three filtersets: (1) near-IR--mid-IR, (2) mid-IR--radio, and (3) near-IR--radio (i.e., all filters).
Case (1) is the same filterset used in \cite{ling24} and case (2) is roughly comparable to \cite{bertoldi07}. We find similar results as previous studies, with Case (1) favoring a high-$z$ solution, although a secondary low-$z$ solution is evident. Case (2) favors a broad single-peaked, low-$z$ solution. Case (3) favors a narrow single-peaked, low-$z$ solution. We note our Case (3) \zphot\ disagrees with \zspec\ by 3$\sigma$, but that it is not a catastrophic failure ($\vert z_\mathrm{phot}-z_\mathrm{spec}\vert/(1+z_\mathrm{spec})=0.06$).

\section{Discussion} 
% of the underlying assumptions being used are reasonable across the datasets to which they are applied. %that it is important to understand whether the underlying assumptions being used are reasonable across the datasets to which they are applied. 
We outline three aspects that can help reduce instances of catastrophic failures:  

First, it is beneficial to use photometry spanning as wide of a wavelength range as possible to break color degeneracies (upper limits also useful). A well-known degeneracy occurs between the Lyman break and Balmer break features, but a similar issue can arise due to the 2175\AA\ absorption feature (e.g., Fig~1). % and which links to our next point.
Sharp SED transitions can also arise due to emission lines \citep[e.g.,][]{zavala23}, which are difficult to self-consistently model with photometry, and wider coverage can mitigate their impact. 
%Testing SED fitting separately on the UV--near-IR and near-IR--sub-mm/radio regions (or excluding particular filters suspected to be impacted by lines) can help to diagnose possible degeneracies in redshift solutions (e.g., Fig~1).

Second, adopting flexible dust attenuation prescriptions that include a varying 2175\AA\ feature can provide more reliable \zphot\ posteriors than a fixed attenuation curve. The average shape of dust attenuation curves likely evolve with redshift \citep[e.g.,][]{battisti22} and % and vary substantially from galaxy to galaxy \citep{salim&narayanan20}. 
there is evidence that a moderate 2175\AA\ feature is typical in SFGs up to $z\sim3$ \citep[e.g.,][]{battisti20}, and has also been seen up to $z\sim7$ \citep[e.g.,][]{witstok23}. 
 
Third, adopting uncertainty floors is important for numerous reasons: (1) formal uncertainties are often underestimated, (2) zeropoint variations between instruments are common, (3) template models used in SED-fitting codes are not perfect representations of reality. The last point is often forgotten and uncertainty floors help account for both limitations in spectral models and the photometry itself. 

Finally, we note COSBO-7 shows a sharp SED transition between F1800W and MIPS24, presumably due to PAH emission. %This aids \zphot\ constraints (see \citealt{jin24}) 
Including MIRI filters into our Case (2) fit gives a tighter constraint of $z_\mathrm{phot}=2.93^{+0.09}_{-0.20}$ from \magphysz\ (not shown), highlighting that the templates describing PAH features impact the results.  
%The relative strength of PAH features to total IR luminosity in SFGs appear relatively stable from $0<z\lesssim2.5$ \citep[e.g.,][]{elbaz11, kirkpatrick15, shivaei24}, however constraints at higher redshifts will remain unavailable until future far-IR missions \citep[e.g., PRIMA,][]{moullet23}. 
For this reason, using flexible IR templates \citep[e.g.,][]{draine&li07} may improve \zphot\ accuracy and reduce potential bias.\\

%The 2175\AA\ feature did not appear to be included in most codes used in \cite{ling24} and this may have resulted in high-$z$ solutions being preferred.

The JWST data presented in this article were obtained from the Mikulski Archive for Space Telescopes (MAST) at the Space Telescope Science Institute. The specific observations analyzed can be accessed via \dataset[doi: 10.17909/qb2s-1t93]{https://doi.org/10.17909/qb2s-1t93}.

%% Please use the acknowledgment and contribution environments. This will 
%% be anonomyized when the "anonymous" style option is used. 
\begin{acknowledgments}
AJB thanks J. Zavala for feedback. AJB and EdC acknowledge funding by Australian Research Council Discovery Project DP240100589. SJ acknowledges Villum Fonden research grants 37440 and 13160. JAH acknowledges support from ERC Consolidator Grant 101088676 (“VOYAJ”).
\end{acknowledgments}

\bibliography{AJB_bib}{}
\bibliographystyle{aasjournalv7}



\end{document}